\documentclass{article} \usepackage{fortschritte} \usepackage{amsmath}
\usepackage{amssymb}
\def\Bbb{\mathbb} \def\C{{\Bbb C}} \def\R{{\Bbb R}} \def\Z{{\Bbb Z}}

    \def\tr{\operatorname{tr}}

 \def\tr{{\rm tr\, }}

\def\id{\protect{{1 \kern-.28em {\rm l}}}}

\newcommand{\be}{\begin{equation}} \newcommand{\ee}{\end{equation}}
\newcommand{\bea}{\begin{eqnarray}} \newcommand{\eea}{\end{eqnarray}}
\newcommand{\beann}{\begin{eqnarray*}}
  \newcommand{\eeann}{\end{eqnarray*}}
\newcommand{\bfig}{\begin{figure}} \newcommand{\efig}{\end{figure}}

\newcommand{\ba}{\begin{array}}\newcommand{\ea}{\end{array}}

\newtheorem{Proposition}{Proposition}[section]

\newtheorem{Theorem}{Theorem}[section]
\newtheorem{Lemma}{Lemma}[section]
\newtheorem{Corrolary}{Corrolary}[section]

\newcommand{\bp}{\begin{Proposition}}
  \newcommand{\ep}{\end{Proposition}}
\newcommand{\bt}{\begin{Theorem}} \newcommand{\et}{\end{Theorem}}
\newcommand{\bl}{\begin{Lemma}} \newcommand{\el}{\end{Lemma}}
\newcommand{\bc}{\begin{Corrolary}} \newcommand{\ec}{\end{Corrolary}}

  \def\om{\omega}
   \def\ep{\eps}
 \def\hn{{ \hat N}} \def\hq{{\hat Q}} \def\hp{{\hat
    \Phi}} \def\hs{{ \hat S}} \def\ha{{\hat A}} 
\def\cn{{\cal N}}

\begin {document}                 
\def\titleline{
%%%%%%%%%%%%%%%%%%%%%%%%%%%%%%%%%%%%%%%%%%%%%%
%                                            
% Insert now the text of your title.         
% Make a linebreak in the title with         
%                                            
%            \newtitleline                   
%                                            
%%%%%%%%%%%%%%%%%%%%%%%%%%%%%%%%%%%%%%%%%%%%%%%
  Puzzles for Matrix Models of Chiral Field Theories
%%%%%%%%%%%%%%%%%%%%%%%%%%%%%%%%%%%%%%%%%%%%%%% 
}
\def\email_speaker{ {\tt
%%%%%%%%%%%%%%%%%%%%%%%%%%%%%%%%%%%%%%%%%%%%%%
%                                                  
% Insert now the e-mail address of the speaker or  
% the author that should get the electronic mail   
% of the publishing house                           
%                                                  
%%%%%%%%%%%%%%%%%%%%%%%%%%%%%%%%%%%%%%%%%%%%%%         
    Karl.Landsteiner@uam.es
%%%%%%%%%%%%%%%%%%%%%%%%%%%%%%%%%%%%%%%%%%%%%%         
  }} \def\authors{
%                                            
%  Insert now the name (names) of the author 
%  (authors).                                
%  In the case of several authors with       
%  different addresses use labels e.g.       
%                                            
%             \1ad  , \2ad  etc.             
%                                            
%  to indicate that a given author has the   
%  address number 1 , 2 , etc.               
%  Signify the person with the above e-mail  
%  address with \sp (behind the address      
%  labels, e.g.                              
%                                            
%            \2ad\sp                         
%                                            
%  when the 2nd author                       
%  happens to be the 'speaker')                                  
%%%%%%%%%%%%%%%%%%%%%%%%%%%%%%%%%%%%%%%%%%%%%%         
  K. Landsteiner \1ad\sp, C. I. Lazaroiu \2ad, R.  Tatar \3ad
%%%%%%%%%%%%%%%%%%%%%%%%%%%%%%%%%%%%%%%%%%%%%%         
} \def\addresses{
%%%%%%%%%%%%%%%%%%%%%%%%%%%%%%%%%%%%%%%%%%%%%%
%                                            
% List now the address. In the case of       
% several addresses list them by numbers     
% using e.g.                                 
%                                            
%             \1ad , \2ad   etc.             
%                                            
% to numerate address 1 , 2 , etc.           
%                                            
%%%%%%%%%%%%%%%%%%%%%%%%%%%%%%%%%%%%%%%%%%%%%
  \1ad
  Instituto de F\'\i sica Te\'orica, Universidad Aut\'onoma de Madrid \\
  28049 Madrid, Spain \\
  \2ad
  5243 Chamberlin Hall, Univ. of Wisconsin, 1150 University Ave\\
  Madison, Wisconsin 53706, USA\\
  \3ad
  Theoretical Physics Group, Lawrence Berkeley National Laboratory \\
  Berkeley, CA 94720, USA
%%%%%%%%%%%%%%%%%%%%%%%%%%%%%%%%%%%%%%%%%%%%%
} \def\abstracttext{We summarize the field-theory/matrix model
  correspondence for a chiral $\cn= 1$ model with matter in the adjoint, 
  antisymmetric and conjugate symmetric representations as well as eight 
  fundamentals to cancel the chiral anomaly. The associated holomorphic matrix model is
  consistent only for two fundamental fields, which requires a modification of the original 
  Dijkgraaf-Vafa conjecture. The modified correspondence holds in spite of this mismatch.}  

%%%%%%%%%%%%%%%%%%%%%%%%%%%%%%%%%%%%%%%%%%%%%%%%%%%%%%%%%%%%%%
%
% Take out Preprint numbers before submitting to Proceedings
%
%
\vspace*{-2truecm} 
\begin{flushright} \tt
IFT-UAM/CSIC-03-44\\MAD-TH-03-3\\LBNL-54038
\end{flushright}
%%%%%%%%%%%%%%%%%%%%%%%%%%%%%%%%%%%%%%%%%%%%%%%%%%%%%%%%%%%%%%
%%%%%%%%%%%%%%%%%%%%%%%%%%%%%%%%%%%%%%%%%%%%%%%%%%%%%%%%%%%%%%

\large \makefront
%%%%%%%%%%%%%%%%%%%%%%%%%%%%%%%%%%%%%%%%%%%%%%%%
%                                              %
%  Insert now the remaining parts of           %
%  your article.                               %
%                                              %
%%%%%%%%%%%%%%%%%%%%%%%%%%%%%%%%%%%%%%%%%%%%%%%%
\section{Introduction}

A new method for determining the effective gaugino
superpotential of confining $\cn =1$ gauge theories was
introduced by Dijkgraaf and Vafa \cite{DV}.  Based on string theory
considerations, they proposed to take the tree-level superpotential as
the ´action´ of a dual holomorphic matrix model \cite{holo}. 
The conjecture was proved for a
few nontrivial examples, via two distinct methods.  One approach
\cite{dglvz} uses covariant superfield techniques in perturbation
theory to integrate out massive matter fields in a gaugino background.
A different method was proposed in \cite{Gorski, cdsw}, where it was
shown that the loop equations of the matrix model coincide formally
with chiral ring relations induced by certain generalizations of the
Konishi anomaly.

We will discuss a model with gauge group $U(N)$ and chiral matter
content chosen to allow for a straightforward large $N$ limit
\cite{chiral1}.  What we mean by this is that the number of chiral
superfields is independent of the rank of the gauge group despite
constraints from anomaly cancellation.  The matter content is given by
a field $\Phi$ in the adjoint representation, two fields $A$, $S$ in
the antisymmetric and conjugate symmetric two-tensor representations,
and eight fields $Q_1\dots Q_8$ in the fundamental representation to
cancel the chiral anomaly.  Because of the chiral spectrum, 
mass terms are not allowed so the methods of
\cite{dglvz} are not applicable.  We chose a tree level
superpotential:
\begin{equation}
  \label{eq:superpot}
  W_{\mathrm{tree}} = \tr [ W(\Phi) + S \Phi A] + 
\sum_{f=1}^{8} Q_f^T S Q_f~~, 
\end{equation}
where $W=\sum_{j}{\frac{t_j}{j}z^j}$ is a complex polynomial of 
degree $d+1$.

\section{The Generalized Konishi Anomalies}
\label{konishis}

Let us discuss the generalized Konishi anomalies for this model, following 
\cite{Gorski} and \cite{cdsw}. The structure of the tree level
superpotential implies:
\begin{equation}
\label{Wj}
  j \frac{\partial W_{eff}}{\partial t_j} = \langle \tr(\Phi^j)\rangle~~.
\end{equation}
The strategy is to extract a set of chiral ring relations which
determine the generating function $T(z) = \langle
\tr(\frac{1}{z-\Phi})\rangle$ of the correlators $\langle
\tr(\Phi^j)\rangle$.  Then integration
of (\ref{Wj}) allows one to compute the effective superpotential up to a
piece independent of $t_j$.

The generalized Konishi anomaly is the anomalous Ward identity for a
local holomorphic transformation of the chiral superfields:
\begin{equation}
{\cal O}^{(r)} \longrightarrow {\cal O}^{(r)} + \delta {\cal O}^{(r)}~~.
\end{equation}
We will project onto the chiral ring, i.e. the equivalence
class of operators which are annihilated by the anti-chiral supercharge
$\bar Q_{\dot\alpha}$. As is well-known, these are in one-to-one
correspondence with the lowest components of chiral superfields.  The
chiral ring relation induced by the generalized Konishi anomaly is:
\begin{equation}
\label{konishi_relation}
 \delta {\cal O}_I
\frac{\partial W}{\partial {\cal O}_I}  \equiv
- \frac{1}{32\pi^2}  {\cal W}^{\alpha}_I\,^J {\cal W}_{\alpha,J}\,^K
\frac{\partial (\delta {\cal O}_K)}{\partial {\cal O}_I }~~,
\end{equation}
where the capital indices enumerate a basis of the representation $r$.
We will investigate the generalized Konishi relations corresponding to
the field transformations:
\begin{eqnarray}
\delta \Phi = \frac{{\cal W}^\alpha {\cal W}_\alpha}{z-\Phi} &,&   
\label{var.phi.one}
\delta \Phi = \frac{1}{z-\Phi}~~, \label{var.phi.two}\\
\delta A = \frac{{\cal W}^\alpha}{z-\Phi}A  
\frac{({\cal W}_\alpha)^T}{z-\Phi^T}  \label{var.A.one}&,&
\delta A = \frac{1}{z-\Phi}A \frac{1}{z-\Phi^T}~~,  \label{var.A.two}\\
\delta S = \frac{1}{z-\Phi^T}S \frac{1}{z-\Phi}  \label{var.S.one} &,&
\delta Q_f = \sum_{g=1}^{N_F}{\frac{\lambda_{fg}}{z-\Phi} Q_g}~~.
\label{var.Q}
\end{eqnarray}
In the last equation, $\lambda$ is an arbitrary matrix in flavor
space.

Writing ${\cal W}^2 = {\cal W}^\alpha {\cal W}_\alpha$, we define:
\begin{eqnarray}
R(z) := - \frac{1}{32\pi^2} \left\langle\tr \left(\frac{{\cal W}^2}{z-\Phi}\right)\right\rangle\label{R.konishi}&,&
T(z) := \left\langle\tr \left(\frac{1}{z-\Phi}\right)\right\rangle~~,\label{T.konishi}\\
M(z) := \left\langle \tr \left( S \frac{1}{z-\Phi}A\right)
\right\rangle\label{M.konishi}&,&
M_Q(z) := \sum_f \left\langle Q_f^T\frac{1}{z-\Phi^T}S\frac{1}{z-\Phi}Q_f \right\rangle~~,
\label{MQ.konishi}\\ 
K(z) := -\frac{1}{32\pi^2}\left\langle \tr
\left( S \frac{{\cal W}^2}{z-\Phi}A\right)
\right\rangle\label{K.konishi} &,&
L(z) := \sum_f \left\langle Q_f^T
S\frac{1}{z-\Phi}Q_f \right\rangle~~,
\label{L.konishi}
\end{eqnarray}
and introduce the degree $d-1$ polynomials:
\begin{eqnarray}
f(z) &:=& -\frac{1}{32\pi^2} \left\langle\tr\left( \frac{W'(z)-W'(\Phi)}{z-\Phi} {\cal W}^2 \right)\right \rangle
\label{fpoly.konishi}~~,\\
c(z) &:=&  \left\langle\tr\left( \frac{W'(z)-W'(\Phi)}{z-\Phi}  \right)\right\rangle \label{cpoly.konishi}~~.
\end{eqnarray}
Assuming that vevs of spinor fields vanish due to Lorentz invariance,
we find the following Ward identities for the generating functions
(\ref{R.konishi}-\ref{L.konishi}):
\begin{eqnarray}
R(z)^2 - 2 W'(z) R(z) + 2 f(z) & = & 0 \label{konishi.R}~~,\\
T(z) R(z) - W'(z) T(z) - 2 R'(z) +  c(z) & = &0 \label{konishi.T}~~,\\
K(z) - \frac{1}{2} R(z)^2 &=& 0 \label{wi.konishi.K}~~,\\
M(z) - R(z) T(z) - 2 R'(z)  &=& 0 \label{wi.konishi.M}~~,\\
M_Q(z) + M(z)  - R(z) T(z) + 2 R'(z)  &=& 0 \label{wi.konishi.MQ}~~,\\
2L(z)- N_F R(z) &=& 0 \label{wi.konishi.L}~~. 
\end{eqnarray}
Given a solution $(R(z), T(z))$ of these constraints, the quantities
$K, M, M_Q$ and $L$ are determined by $R(z)$ and $T(z)$. Hence all
solutions are parameterized by the $2d$ complex coefficients of the
polynomials $f(z)$ and $c(z)$.

The generalized Konishi relations involving the flavors $Q_f$ have an
interesting implication. Expanding equations (\ref{wi.konishi.K})
and (\ref{wi.konishi.L}) to leading order in $1/z$ gives:
\begin{eqnarray}
S=\frac{1}{4}\sum_f \langle Q_f^T S Q_f \rangle =
\frac{2}{N_F}\sum_f \langle Q_f^T S Q_f \rangle~~,
\end{eqnarray}
where $S=-\frac{1}{32\pi^2} \langle \tr {\cal W}^2\rangle$ is the
gaugino condensate. If $S$ is non-vanishing, then compatibility of
these two equalities requires that we set $N_F=8$, which is also
necessary to cancel the chiral anomaly.  Any other value is
incompatible with the existence of a gaugino condensate.

\section{The Matrix Model}
\label{matrixmodel}
The general conjecture of \cite{DV} suggests that the effective
superpotential of our field theory should be described by the
holomorphic \cite{holo} matrix model:
\begin{equation}
\label{Z_mod}
Z=\frac{1}{|G|}\int_\Gamma{d\mu e^{-\frac{\hn}{g}
{\cal S}_{mm}({\hat \Phi}, {\hat A}, {\hat S}, {\hat Q})}}~~,
\end{equation}
where $|G|$ is a normalization factor, $\Gamma$ denotes a gauge
equivalence class of paths in the complex matrix configurations space
$\cal M$ with $\mathrm{dim}_\R(\Gamma) =\mathrm{dim}_\C({\cal M})$ and
the matrix model action is given by:
\begin{equation}
\label{action}
{\cal S}_{mm}({\hat \Phi}, {\hat A}, {\hat S}, {\hat Q})=
\tr\left[ W({\hat \Phi})+{\hat S}{\hat \Phi} {\hat A} \right]+\sum_{f=1}^{{\hat N}_F}{{\hat Q}_f^T {\hat S} {\hat Q}_f}=
\tr\left[ W({\hat \Phi})+{\hat \Phi} {\hat A} {\hat S}+\sum_{f=1}^{{\hat N}_F}{{\hat Q}_f {\hat Q}_f^T {\hat S}} \right]~~.
\end{equation}
We use the convention that all matrix model quantities are denoted by
hatted capital letters.  The measure
$$
d\mu = \bigwedge_{i,j}^{\hat N} d\hat \Phi_{ij}
\bigwedge_{i< j =1}^{\hat N} d\hat A_{ij} \bigwedge_{i\leq j
=1}^{\hat
N} d\hat S_{ij} \bigwedge_{f=1}^{\hat N_f}\bigwedge_{i=1}^{\hat N}
d\hat
Q_{i,f}
%d\mu = \bigwedge_{i< j =1}^N d\hat A_{ij} \bigwedge_{i\leq j
%=1}^N
%d\hat S_{ij} \bigwedge_{f=1}^{N_f}\bigwedge_{i=1}^{N} d\hat Q_{i,f}
$$
is a top holomorphic form on the space of matrices.  Since anomaly
cancellation in our field theory requires $N_F=8$, one is tempted to
set ${\hat N}_F=8$ as well. Note that we are forced to use a purely holomorphic formulation as in 
\cite{holo}, since one cannot impose a reality condition on the matrices 
(such as hermiticity) without breaking the gauge invariance of the matrix model.

It turns out that the naive
identification (\ref{Z_mod}) cannot hold in our case.  The matrix model action is
invariant under the (complexified) gauge group $GL({\hat N},\C)$
acting as:
\begin{equation}
  \label{matrix_gauge} \hat \Phi \rightarrow U \hat \Phi U^{-1} ~~~,~~~
  \hat A\rightarrow U \hat A U^T~~~,~~~\hat S\rightarrow (U^{-1})^T
\hat S U^{-1}~~~,~~~ \hat Q_f \rightarrow U \hat Q_f\,.
\end{equation}
However, the measure $d\mu$ is {\em not} invariant under the central
$\C^*$ subgroup of $GL({\hat N},\C)$ unless ${\hat N}_F=2$. Taking
$U=\xi {\bf 1}_{\hat N}$ with $\xi\in \C^*$, we have:
\begin{equation}
\label{U1}
{\hat A}\rightarrow \xi^2 {\hat A}~~,~~{\hat S}\rightarrow
\xi^{-2}{\hat S}~~{\rm and}~~
{\hat Q}_f\rightarrow \xi {\hat Q}_f~~,
\end{equation}
which gives:
\begin{equation} Z=\xi^{{\hat N}({\hat
    N}_F-2)} Z~~.  
\end{equation} 
Thus $Z$ must either vanish or equal complex infinity! This means that
the matrix model predicted by a naive application of the conjecture of
\cite{DV} is not well-defined.

That subtleties can arise when attempting to apply the conjecture of
\cite{DV} to chiral field theories is not completely unexpected, since
most derivations of this conjecture up to date have concentrated on
real matter representations, which prevent the appearance of net
chirality\footnote{Note however \cite{Brandhuber} who study
  generalized Konishi anomalies for some chiral models, though 
without studying the associated matrix models.}.  The
phenomenon we just discussed shows that one must modify the original
conjecture of \cite{DV} in order to adapt it to the chiral context.

We also note that the superpotential involves only operators which are
singlets under the flavor group.  Therefore one might still hope to 
find a formal map between the loop equations and the Konishi
anomalies which also involves only flavor singlets.  This leads us to
consider the matrix model with ${\hat N}_F=2$.  Then both the action
(\ref{action}) and the integration measure are invariant under
$GL(\hn,\C)$ transformations of the form (\ref{matrix_gauge}), where
$U$ is now an arbitrary complex invertible matrix.

Let us have a look at the loop equations of the (\ref{Z_mod}).  
Although the correlation functions are not well
defined unless $\hn_F=2$, we shall consider formal expressions with an
arbitrary value of ${\hat N}_F$.  This will allow us to re-discover
the constraint ${\hat N}_F=2$ as a consistency condition between the
loop equations, similar to the way in which we rediscovered
the condition $N_F=8$ by using the Konishi constraints of the field
theory.

We consider the generalized the matrix model resolvents:
\begin{eqnarray}
\label{omega}
\omega(z) = \frac{g}{\hn} \tr \left(\frac{1}{z-\hp}\right) &~~,~~&
k(z) = \frac{g}{\hn}\tr \left( \hs \frac{1}{z-\hp} \ha \right)~~,
\label{def.m}\\
m_Q(z) = \hq_f^T \frac{1}{z-\hp^T} \hs \frac{1}{z-\hp} \hq_f
\label{def.mQ}  &~~,~~&
l(z) = \hq_f^T \hs\frac{1}{z-\hp} \hq_f\label{def.l}~~.
\end{eqnarray}
It is not hard to show that they fulfill the loop equations:
\begin{eqnarray}
\langle \om (z)^2 - \frac{g}{\hn} \om'(z) - 2 W'(z) \om(z )
+ 2 \tilde{f}(z) \rangle &=&0~~,\label{loop.om}\\
\langle \frac 1 2 \om(z)^2  + \frac 1 2 \frac{g}{\hat N} \om'(z)  - k(z)\rangle &=&0~~,\label{loop.m} \\
\langle \om'(z)  +  m_Q(z) \rangle &=&0~~,\label{loop.mq}\\
\langle {\hat N}_F \omega(z)-2l(z) \rangle&=&0~~,\label{loop.l}
\end{eqnarray}
where:
\begin{equation}
\label{tilde_f}
\tilde{f}(z) := \frac{g}{\hn}\tr \frac{W'(z)-W'(\hp)}{z-\hp} ~~
\end{equation}
is a polynomial of degree $d-1$.
Note that the only dynamical input is represented by
the polynomial $\tilde f(z)$.

The leading order in the large $z$ expansion of the last two equations
gives:
\begin{eqnarray}
  \label{eq:matrix.consisitency}
  g= \langle \hq_f^T \hs \hq_f \rangle = 
  \frac{2}{\hn_F}\langle \hq_f^T \hs \hq_f \rangle\,.
\end{eqnarray}
Since we of course take $g\neq 0$, equations
(\ref{eq:matrix.consisitency}) are consistent only if ${\hat N}_F=2$.

\section{Relation between the matrix model and field theory}
\label{matrix.vs.field}

Consider the large ${\hat N}$ expansion of a generic matrix model
expectation value $\langle \hat{X}\rangle$:
\begin{equation}
\langle \hat X \rangle
=\sum_{j=0}^\infty{\left(\frac{g}{\hn}\right)^j \hat X_j}~~.
\label{expansion.omega}
\end{equation}

Applying this expansion to (\ref{loop.om})-(\ref{loop.l}) it is
clear that to leading order we obtain equations which are formally
identical to the Konishi constraints for $R(z), K(z), M_Q(z)$
and $L(z)/4$. This implies of course that we identify the
polynomials $f(z)$ and $\tilde f(z)$.

Taking into account also the terms at order $g/{\hat N}$ and
introducing the operator $\delta:=\sum_{k}{N_k\frac{\partial}{\partial
    S_k}}$, it is easy to see that $\delta \omega_0(z)+4\omega_1(z)$
can be identified with $T(z)$ and $ \delta k_0(z)+4k_1(z)$ with
$M(z)$.

To summarize, we recover the Konishi constraints from the
large $\hat N$ expansion of the matrix model loop equations by
making the following identifications:
\begin{center}
\begin{tabular}{|c|c|}
\hline
Matrix~Model & Field Theory\\ \hline
$\omega_0(z)$ & $R(z)$ \\
$\delta \omega_0(z) + 4 \omega_1(z)$ & $T(z)$\\
$\delta k_0(z) + 4 k_1(z)$ & $M(z)$ \\
$k_0(z)$ & $K(z)$\\
$4m_{Q0}(z)$ & $M_Q(z)$\\
$4 l_0(z)$ & $L(z)$\\
$\tilde f_0(z)$ & $f(z)$\\
$\delta \tilde f_0(z) + 4 \tilde f_1(z)$ & $c(z)$ \\
\hline
\end{tabular}
\end{center}

%\begin{center}
%  {\footnotesize Table 1: Identification between field theory and
%    matrix model quantities.}
%\end{center}

This also implies that the effective superpotential is given by
\begin{equation}
\label{matrix.model.superpot}
W_{eff}(t,S)=\sum_{i}{N_i \frac{\partial F_0}{\partial S_i}}+4 F_1+\psi(S)~~.
\end{equation}
Here $F=\log Z$ is the matrix model's partition function and $\psi$ is
a function which depends on $S_i$ but not on the coefficients of
$W$.  We take $\label{psi} \psi(S)=\alpha \sum_{i=1}^{d}{S_i}~~,$ with
$
\label{alpha}
\alpha=(N-4)\ln \Lambda~~ $, where $\Lambda$ is the scale of our field theory.  
Then one can check that $\psi(S)$ together with the
non-perturbative contribution to $F$ due to the measure of the matrix
model give the Veneziano-Yankielowicz part of
the effective superpotential \cite{chiral1}.

\section{Conclusions}
\label{conclusion}

Rather surprisingly, we found that the number of fundamental flavors
$N_F$ in field theory must be taken to differ from the number of
flavors ${\hat N}_F$ in the dual matrix model.  Nevertheless, there
exists an exact if formal one-to-one map between the large $\hat N$
expansion of the loop equations of the matrix model and the Konishi
anomaly constraints of the field theory.  A better understanding of
this mismatch between the numbers of flavors might  be obtained by
studying the string theory realization of this model.  
The basic analysis was performed in \cite{chiral2},
where we showed that, unexpectedly, the geometric
background involves a $\Z_4$ orientifold of an $A_2$ fibration and
that the flavors correspond to eight fractional D5-branes.

\bigskip
{\bf Acknowledgments:} This work was supported by DFG grant KL1070/2-1 (C.I.L.)
and by DOE Contract DE-AC03-76SF00098 and NSF grant PHY-0098840 (R.T.).
\label{acknowledgements}


\begin{thebibliography}{77}
\bibitem{DV}{R.~Dijkgraaf and C.~Vafa, ``Matrix models, topological
    strings, and supersymmetric gauge theories,'' Nucl.\ Phys.\ B {\bf
      644}, 3 (2002) [arXiv:hep-th/0206255];\\
%%CITATION = HEP-TH 0206255;%%
    R.~Dijkgraaf and C.~Vafa, ``On geometry and matrix models,''
    Nucl.\ Phys.\ B {\bf 644}, 21 (2002)
    [arXiv:hep-th/0207106];\\
%%CITATION = HEP-TH 0207106;%%
    R.~Dijkgraaf and C.~Vafa, ``A perturbative window into
    non-perturbative physics,'' arXiv:hep-th/0208048.}
%%CITATION = HEP-TH 0208048;%%
\bibitem{holo}{C.~I.~Lazaroiu, ``Holomorphic matrix models,'' JHEP {\bf 0305}, 044 (2003) [arXiv:hep-th/0303008].}
%%CITATION = HEP-TH 0303008;%%
\bibitem{dglvz}{ R.~Dijkgraaf, M.~T.~Grisaru, C.~S.~Lam, C.~Vafa and
    D.~Zanon, ``Perturbative computation of glueball
    superpotentials,'' Phys.\ Lett.\ B {\bf 573}, 138 (2003)
    [arXiv:hep-th/0211017].}
\bibitem{Gorski}{A.~Gorsky, ``Konishi anomaly and N = 1 effective
    superpotentials from matrix models,'' Phys.\ Lett.\ B {\bf 554},
    185 (2003) [arXiv:hep-th/0210281].}
%%CITATION = HEP-TH 0210281;%%
\bibitem{cdsw}{F.~Cachazo, M.~R.~Douglas, N.~Seiberg and E.~Witten,
    ``Chiral rings and anomalies in supersymmetric gauge theory,''
    JHEP {\bf 0212}, 071 (2002) [arXiv:hep-th/0211170].}
%%CITATION = HEP-TH 0211170;%%
\bibitem{chiral1} K.~Landsteiner, C.~I.~Lazaroiu and R.~Tatar,
  ``Chiral field theories, Konishi anomalies and matrix models,''
  JHEP {\bf 0402} 044 (2004) [arXiv:hep-th/0307182].
%%CITATION = HEP-TH 0307182;%%
\bibitem{chiral2} K.~Landsteiner, C.~I.~Lazaroiu and R.~Tatar,
  ``Chiral field theories from conifolds,'' JHEP {\bf 0311} 057 (2003)
  [arXiv:hep-th/0310052].
%%CITATION = HEP-TH 0310052;%%
\bibitem{Brandhuber} A.~Brandhuber, H.~Ita, H.~Nieder, Y.~Oz and
  C.~Romelsberger, ``Chiral rings, superpotentials and the vacuum
  structure of N = 1 supersymmetric gauge theories,'' Adv.\ Theor.\ 
  Math.\ Phys.\ {\bf 7}, 269 (2003) [arXiv:hep-th/0303001];\\
%%CITATION = HEP-TH 0303001;%%
R.~Argurio, G.~Ferretti and R.~Heise, ``Chiral
    SU(N) gauge theories and the Konishi anomaly,'' 
    JHEP {\bf 0307} 044 (2003) [arXiv:hep-th/0306125].
%%CITATION = HEP-TH 0306125;%%
\end{thebibliography}
\end{document}